\documentclass[letterpaper,twocolumn,english,prl,showpacs]{revtex4}

\makeatletter
\usepackage{graphicx}
\usepackage{amsthm}
\usepackage{amsfonts}

\newcommand{\Ket}[1]{\ensuremath{\left | #1 \right \rangle}}
\newcommand{\Bra}[1]{\ensuremath{\left \langle #1 \right |}}

\usepackage{babel}
\makeatother
\begin{document}

\title{Nondeterministic testing of Sequential Quantum Logic propositions
on a quantum computer}

\author{M. S. Leifer}

\email{mleifer@perimeterinstitute.ca}

\affiliation{Perimeter Institute for Theoretical Physics, 31 Caroline Street North,
Waterloo, Ontario, Canada, N2L 2Y5}


\begin{abstract}
In the past few years it has been shown that universal quantum computation
can be obtained by projective measurements alone, with no need for
unitary gates. This suggests that the underlying logic of quantum
computing may be an algebra of sequences of quantum measurements rather
than an algebra of products of unitary operators. Such a Sequential Quantum Logic
(SQL) was developed in the late 70's and has more recently been applied
to the consistent histories framework of quantum mechanics as a possible
route to the theory of quantum gravity. In this letter, I give a method
for deciding the truth of a proposition in SQL with nonzero probability
of success on a quantum computer. 
\end{abstract}
\maketitle
Formal logic plays a central role in the theory of classical computation and leads to many powerful results in computability \cite{BoolosBurgessJeffrey}, computational complexity \cite{Papadimitriou} and the theory of programming languages \cite{Reynolds}. On the other hand, quantum computing is not normally presented
in formal terms, and one might wonder whether developing a logic
of quantum computing would lead to similar insights. As a simple example
of the sort of result that might be raised
to the quantum domain, consider any computation in the classical circuit
model that produces a one-bit output, i.e. a circuit for solving a
decision problem. It is clear that such a computation simply evaluates
the truth-value of a Boolean formula, given the truth-values of its
inputs. This simple observation is directly related to the Cook-Levin
theorem, which states that SATISFIABILITY - the problem of determining
whether a Boolean formula has any satisfying assignments - is NP complete,
and indeed that similar logical problems are complete for all complexity
classes in the polynomial hierarchy. In turn, this means that some
of the most important problems in computer science - whether P = NP
and whether the polynomial hierarchy collapses - can be determined
by the equivalence or non-equivalence in complexity of these logical
decision problems.

The Extended Church-Turing thesis, that any
physically reasonable model of computing can be simulated by a probabilistic Turing machine
with only polynomial overhead, implies that such results are robust
against the particular choice of physical realization used
to build a computer. Hence, classical computer scientists do not
have to learn the full details of Newtonian physics to study their
subject. Instead, they can abstract away from the physics by realizing
that ``information is logical''.

The discovery of quantum computing has compelled us to believe
that in fact ``information is physical''. Physically realizable
models of computing may be more powerful than those based
on the abstraction of Boolean logic. This point is forcefully demonstrated
by the existence of quantum algorithms that run exponentially faster
than any known classical algorithm for solving the same problem \cite{Shor}. It
seems that the ability to evaluate polynomial length Boolean logic
formulas does not capture the essence of what is efficiently computable
in our universe, so perhaps information is not merely logical after
all.

However, given the unifying power of logic in classical computer science,
we should be reluctant to give up a logical notion of information
processing altogether. Perhaps the shift from classical to quantum
computing is simply a shift in the logic that underlies our models
of computing. This has been suggested by Deutsch et. al. \cite{DeutschEkertLupMLQP}, who speculated
that reversible classical logic should be expanded to include other
unitary operations that occur in the circuit model of quantum computing,
such as $\sqrt{\text{NOT}}$. These ideas have been formalized by dalla Chiara et. al. \cite{DallaChGiunLepSurvey}, but their relevance to the general theory of quantum computing,
and particularly to quantum complexity theory, remains unclear at present  \footnote{Very different approaches, directed towards the formal semantics side of computer
science, can be found in \cite{Semantics}.}.

The idea that the shift from classical to quantum requires
a shift in logic is not new. It was proposed in a different context
by Birkhoff and von Neumann in 1936 \cite{BvNQL}. They were concerned with the
logic of \emph{properties} of quantum systems, so theirs is a logic
of possible alternative measurements rather than of unitary operators. On the other hand,
quantum computing requires a logic of \emph{processes}, i.e. a description
of how quantum data can be manipulated by sequences of operations.
Classically, Boolean algebra can be viewed as both a logic
of property \emph{and} of process, the latter via the classical circuit model,
but a priori the analogous quantum logics might be much less closely
related.  Nevertheless, the power of quantum computers may reside in an ability to efficiently decide propositions in some sort of quantum logic, more general than the Boolean logic underlying classical computing \footnote{Interestingly, the enhanced power is not captured by the ability to decide propositions in Birkhoff-von Neumann quantum logic.  I have constructed a model of computing capable of doing this in close analogy with the classical circuit model, but it can be simulated efficiently on a classical computer \cite{AaronsonPriv}.}.

The advent of universal models of quantum computation based
on measurements \cite{Meas} suggests that the
most natural logic of quantum computation might be an algebra of sequences of measurements
rather than a logic of unitary operators. Such a Sequential Quantum
Logic (SQL) was developed by Mittlestaedt and Stachow \cite{Mittle, Stachow} and has more
recently been extended by Isham et. al. \cite{IshamQL} as a route
to a quantum theory of gravity within the consistent histories approach
to quantum mechanics. SQL is therefore a natural starting point for
developing a logic of quantum computation based on measurements.

In this letter, we give a method for evaluating SQL propositions on
a quantum computer with non-zero probability of success. It is inspired by recent applications of quantum information ideas to the Density
Matrix Renormalization Group (DMRG) that is used extensively for numerical
simulations of many-body systems \cite{DMRG}. We begin with a
review of SQL, followed by a simple example of the algorithm to illustrate the general procedure and conclude with open problems
suggested by this work.

The syntax of SQL is given by $\langle L,S,\sqcap,\lnot,(,)\rangle$,
where $L$ is a set of elementary propositions and $S$ is the set of sequential propositions.  Propositions in $L$ are denoted $a,b,c,\ldots$ to distinguish them from generic propositions in $S$, denoted $s,t,u,\ldots$.  $S$ is defined recursively by: 

\begin{itemize}
\item If $a\in L$ then $a\in S$. 
\item If $s,t\in S$ then $(s\sqcap t)\in S$. 
\item If $s\in S$ then $\lnot s\in S$. 
\end{itemize}
The aim of SQL is to model sequences of two-outcome measurements,
or \emph{tests}, made on a quantum system. The operators $\lnot,\sqcap$
have the interpretation of negation and sequential conjunction, or
``not'' and ``and then'' respectively. An elementary test is associated
with a pair $\{ a,\lnot a\}$, representing the two possible outcomes. If the system is then subjected to another test represented
by $\{ b,\lnot b\}$ then the four outcomes are represented
via the $\sqcap$ operator as $\{ a\sqcap b,a\sqcap\lnot b,\lnot a\sqcap b,\lnot a\sqcap\lnot b\}$.
Sequential conjunction differs from a regular conjunction in that
$a\sqcap b$ does not necessarily mean that $a$ and $b$ are ever both true at the same time. $\sqcap$ is
not assumed to be commutative, $s\sqcap t\neq t\sqcap s$, but is
associative, $(s\sqcap t)\sqcap u=s\sqcap(t\sqcap u)=s\sqcap t\sqcap u$,
which facilitates the removal of brackets from long expressions. Full
details of the syntax of SQL can be found in \cite{Stachow}, but the
above is sufficient for our present purpose.

SQL is modeled by an algebra of operators acting on a Hilbert space
$\mathcal{H}$.  Given a proposition $s$, denote by $[s]$ the operator on $\mathcal{H}$
assigned to $s$.  Throughout, we assume that states are updated according to the L{\"u}ders-von Neuman projection postulate after a measurement and states are left unnormalized. For the elementary propositions, recall that in quantum mechanics, elementary measurements are represented by sets of projection operators $\{P_j\}$, where $\sum_j P_j = I$ and $I$ is the identity operator.  If the state of the system prior to the measurement is $\Ket{\Psi}$ then, upon obtaining outcome $P_n$, the post-measurement state is $P_n \Ket{\Psi}$.  Therefore, it is natural to represent an elementary proposition $a$ by a projection operator $[a] = P$.  The other outcome of the test, $\lnot a$, is represented by $[\lnot a] = I - [a] = I - P$.  Given a state $\Ket{\Psi} \in \mathcal{H}$, the elementary proposition $a$ can be tested by performing the measurement corresponding to the projectors $[a]$ and $[\lnot a]$, obtaining either the state $[a]\Ket{\Psi}$ or the state $[\lnot a] \Ket{\Psi}$ with probabilities given by the norm squared of the states.  

Having determined the Hilbert space representation of elementary propositions, the operator $[a \sqcap b]$ can be defined.  Consider making a sequence of two measurements on $\Ket{\Psi}$ corresponding to sets of projectors $\{P_j\}$ and $\{Q_k\}$ respectively.  Upon obtaining the outcome $P_n$ of the first measurement followed by the outcome $Q_m$ of the second, the state is updated to $Q_m P_n \Ket{\Psi}$.  Thus, it is natural to set $[a \sqcap b] = [b][a]$.  To complete the algebra, the definitions are extended to all sequential propositions, i.e. $[\lnot s] = I - [s]$ and $[s \sqcap t] = [t][s]$.  For later convenience, the notation $[s^0] = [\lnot s], [s^1] = [s]$ is also used.

Sequences of two-outcome measurements are described correctly by SQL,
provided the measurement results of the entire sequence are retained.
For example if two elementary propositions, $a$ and $b$, are tested
sequentially on a state $\Ket{\Psi}$, then the four states $[a\sqcap b]\Ket{\Psi},[a\sqcap\lnot b]\Ket{\Psi},[\lnot a\sqcap b]\Ket{\Psi}$
and $[\lnot a\sqcap\lnot b]\Ket{\Psi}$ are obtained after the measurement with
probabilities given by the norm squared of the states. However, coarse grainings of such sequences
are not related to physical tests in such a simple manner. For example,
the pair $\{ a\sqcap b,\lnot(a\sqcap b)\}$ is generally not a possible
two-outcome test in quantum mechanics. To see this, note that according to the Hilbert space
model of SQL, the second outcome is associated with the operator $I-[b][a]$,
which is the sum of the three operators $[\lnot a\sqcap b],[a\sqcap\lnot b]$
and $[\lnot a\sqcap\lnot b]$ that appear in the fine grained
version of the test described above. This means that the negative
outcome of the coarse-grained test should result in a coherent superposition
of the states $[a\sqcap\lnot b]\Ket{\Psi},[\lnot a\sqcap b]\Ket{\Psi}$
and $[\lnot a\sqcap\lnot b]\Ket{\Psi}$ rather than the incoherent
mixture that would be obtained by performing the fine grained
test and discarding information about some of the outcomes. In fact, the
operators $\{[b][a],(I-[b][a])\}$ are not a pair of generalized measurement
operators, since $([b][a])^{\dagger}([b][a])+(I-[b][a])^{\dagger}(I-[b][a])\neq I$
unless $[a]$ and $[b]$ commute, so there is no direct quantum mechanical
implementation of the test $\{ a\sqcap b,\lnot(a\sqcap b)\}$. %
%\footnote{The same issue arises in the Consistent Histories formalism, where
%a pair such as $\{ a\sqcap b,\lnot(a\sqcap b)\}$ is called an \emph{inhomogeneous} 
%family of histories \cite{IshamQL}. However, in that approach
%histories are not interpreted as sequences of measurements that are
%actually performed, but are in some sense ``realized'' as possible
%alternatives when no measurement occurs. %Whether this is a meaningful
%notion is a subject of considerable debate. In the approach of Isham
%et. al., inhomogeneous histories are avoided by formally replacing
%the product of operators $[b][a]$ with the tensor product $[b]\otimes[a]$
%on $\mathcal{H}\otimes\mathcal{H}$ \cite{IshamQLTalk, IshamQL}. Elements of $S$ are then always
%represented by projection operators. However, to accommodate this
%and retain the probabilities predicted by quantum mechanics, the Born
%rule has to be modified \cite{IshamLindenSchrek}, so this does not help with the physical implementation
%of the test.}. 

Another way of seeing this is to note that although $\Ket{\psi}=[s]\Ket{\psi}+[\lnot s]\Ket{\psi}$ for any $s$ and $\Ket{\psi}$, the states $[s]\Ket{\psi}$ and $[\lnot s]\Ket{\psi}$ are generally not orthogonal so they cannot be distinguished with certainty. However, It is possible to perform the test nondeterministically by introducing
a third ``failure'' outcome, as in the unambiguous discrimination of nonorthogonal states \cite{Chefles}. Conditional on not obtaining
this outcome, the state $[s]\Ket{\psi}$ is obtained with probability
$N \Bra{\psi}[s]\Ket{\psi}$, and the state $[\lnot s]\Ket{\psi}$ is
obtained with probability $N \Bra{\psi}[\lnot s]\Ket{\psi}$, where $N$ is a positive constant.  This is what is meant by the nondeterministic testing of a proposition $s$ and we now describe an algorithm that achieves this by means of a simple example.

Suppose we want to test the SQL proposition $s=\lnot(a\sqcap b)\sqcap c$.
For brevity, $x$ will be used to denote a generic elementary proposition,
i.e. it should always be understood that $x=a,b,c$. Suppose that
each elementary proposition $x$, is assigned to a projector $[x]=\Ket{\psi_{x}}\Bra{\psi_{x}}$ onto a state $\Ket{\psi_{x}}$ on $\mathbb{C}^{2}$; the Hilbert space
of a single qubit. Let $U_{x}$ be the unitary operator that acts
as $U_{x}\Ket{1}=\Ket{\psi_{x}}$, $U_{x}\Ket{0}=\Ket{\psi_{\lnot x}}$
where $\Ket{\psi_{\lnot x}}$ is any representative state in the subspace
orthogonal to $\Ket{\psi_{x}}$. Thus, for each $x$, we have $[x^{0}]=[\lnot x]=U_{x}\Ket{0}\Bra{0}U_{x}^{\dagger}$
and $[x^{1}]=[x]=U_{x}\Ket{1}\Bra{1}U_{x}^{\dagger}$.

Throughout the algorithm, ancillary qubits are used to coherently store the results of testing sub-propositions of $s$, i.e. $\{a,b,c\}$ at the beginning, followed by $\{a\sqcap b,c\}$, followed by $\{\lnot(a\sqcap b),c\}$, and finally $s$.  It is convenient to use these sub-propositions as labels for the ancillas and to label the qubit that the output state ends up in by $f$.  The procedure to test $s$ on a state $\Ket{\Psi}$ consists of two stages.   The first stage is to prepare the state 
\begin{equation}
\label{history}
\sum_{j,k,m=0}^{1}\Ket{j}_{a}\Ket{k}_{b}\Ket{m}_{c}[c^{m}][b^{k}][a^{j}]\Ket{\Psi}_{f},
\end{equation}
 which we call a \emph{history state}, since, in the computational
basis, the qubits $a,b$ and $c$ encode which of the $2^{3}$
possible sequences of measurement outcomes have occurred. In other words,
the $a,b$ and $c$ qubits contain full information about the outcomes
of a sequence of tests: $\{ a,\lnot a\}$ followed by $\{ b,\lnot b\}$
followed by $\{ c,\lnot c\}$.  The information we require about $s$
is encoded in the correlations between these qubits,
and is obtained in the second stage by two rounds of DMRG-like operations, which compute history states for the sub-propositions of $s$, reducing the dimensionality of the Hilbert space until there is only a single qubit left which stores the result of the test $\{ s,\lnot s\}$. 

The preparation stage begins with the state
\begin{equation}
\label{start}
\Ket{\Psi}_{a}\Ket{\Phi^{+}}_{a' b}\Ket{\Phi^{+}}_{b' c}\Ket{\Phi^{+}}_{c' f},
\end{equation}
 where $\Ket{\Phi^{+}}= \Ket{00}+\Ket{11}$ and $a', b', c'$ are additional ancillas.  By performing measurements followed by a unitary correction, this can be transformed into (\ref{history}). The first step is
to perform a local unitary operation $U_{a}^{\dagger}\otimes U_{a}^{T}$
on qubits $a$ and $a'$, where $^{T}$ is the transpose in the computational
basis. Then a parity measurement is performed on $a$ and $a'$, with
outcomes corresponding to the projectors $P_{1}=\Ket{00}\Bra{00}_{aa'}+\Ket{11}\Bra{11}_{aa'}$
and $P_{2}=\Ket{01}\Bra{01}_{a a'}+\Ket{10}\Bra{10}_{a a'}$.
If outcome $P_{2}$ occurs, then a unitary correction $U_{a}XU_{a}^{\dagger}$
is applied to the qubit $b$, where $X=\Ket{0}\Bra{1}+\Ket{1}\Bra{0}$
is the bit-flip operator. In both cases a CNOT operation is performed
with $a$ as the control and $a'$ as the target. This disentangles
$a'$ from the other qubits and it can then be discarded. The same procedure is then applied to the $b,b'$  and $c,c'$ qubits, where, in the case of $c,c'$, the correction is applied to the qubit $f$ if $P_2$ occurs.  This results in the state (\ref{history}) \footnote{The preparation of the history state is the only part of the algorithm that changes if the restriction to rank 1 projectors onto qubit states is removed.  If, for each pair $[x],[\lnot x]$, either they are of equal rank or one of them is the identity and the other is the null projector, then there is a similar teleportation-like procedure to prepare the history state deterministically starting with (\ref{start}), with $\Ket{\Phi^+}$ a maximally entangled state of appropriate dimension.  If this is not the case then, since the history state just coherently records the outcomes of three successive measurements, it can be prepared deterministically without (\ref{start}) by starting from the state $\Ket{\Psi}_f$, introducing 3 ancilla qubits in the state $\Ket{0}$ and performing the unitary operations $\sum_{jk=0}^{1} \Ket{j}_x\Bra{k}_x \otimes [x^{j \oplus k}]$.}.

In the the second stage, the first step is to compute $a\sqcap b$.  This could be done if it were possible to apply a coherent AND gate, $A_{a,b}=\Ket{0}_{a\sqcap b}(\Bra{00}+\Bra{01}+\Bra{10})_{ab}+\Ket{1}_{a\sqcap b}\Bra{11}_{ab}$, to (\ref{history}), resulting in the state $\sum_{j,k=0}^{1}\Ket{j}_{a\sqcap b}\Ket{k}_{c}[c^{j}][a\sqcap b]\Ket{\Psi}_{f}$.  This is a history state encoding the results of the sub-formulae
$a\sqcap b$ and $c$. Unfortunately, it is not possible to implement $A_{a,b}$ directly, since
the largest eigenvalue of $A_{a,b}^{\dagger}A_{a,b}$ is greater than $1$.  Instead,
a measurement can be performed that has two generalized measurement
operators given by \begin{equation}
\begin{array}{lll}
M_{a,b}^{(s)} & = & \frac{1}{\sqrt{3}}\left(\Ket{01}_{ab}\left(\Bra{00}+\Bra{01}+\Bra{10}\right)_{ab}\right.\\
 &  & \left.\qquad\qquad\qquad\qquad+\Ket{11}_{ab}\Bra{11}_{ab}\right)\\
M_{a,b}^{(f)} & = & (I-M_{s}^{\dagger}M_{s})^{1/2},\end{array}\end{equation}
 wherein obtaining the $M^{(s)}$ outcome, discarding the qubit $b$
and relabeling $a$ as $a\sqcap b$ gives the desired result. Since
we know that SQL propositions cannot be tested deterministically,
and the coherent AND is the only probabilistic stage of this algorithm,
it cannot be possible to perform a correction if the $M^{(f)}$ outcome
is obtained in general. Therefore, on obtaining $M^{(f)}$, the whole
procedure has to be repeated from the beginning until a successful
outcome is obtained.

The next step is to perform a bit-flip operation $\Ket{0}_{\lnot(a\sqcap b)}\Bra{1}_{a\sqcap b}+\Ket{1}_{\lnot(a\sqcap b)}\Bra{0}_{a\sqcap b}$
in order to negate the outcome of the test of $a\sqcap b$. This results
in the state $\sum_{j,k=0}^{1}\Ket{j}_{\lnot(a\sqcap b)}\Ket{k}_{c}[c^{j}][\lnot(a\sqcap b)^{j}]\Ket{\Psi}_{f}$.

Finally, we need to compute the sequential AND of $\lnot(a\sqcap b)$
and $c$. This is done by applying the operator $A_{\lnot(a\sqcap b),c}$,
using the probabilistic procedure described above. This results in
the state $\sum_{j=0}^{1}\Ket{j}_{s}[s^{j}]\Ket{\Psi}_{f}$,
 where $s = \lnot (a \sqcap b) \sqcap c$.  Measuring the qubit $s$ in the computational basis performs the
test $\{ s,\lnot s\}$ on the state $\Ket{\Psi}_{f}$, which is the
desired result.

The generalization to more complicated propositions is straightforward, and this gives the full algorithm for deciding SQL propositions.  This is analogous to solving a Boolean decision problem via a classical circuit with a single output.  To obtain the full classical circuit model, one has to consider circuits with multiple outputs, which correspond to testing multiple Boolean propositions on the same inputs simultaneously.  Likewise, to build a full model of computing based on SQL, one would need to consider how multiple propositions might be tested in the same computation.  One way of doing this  would be to use a quantum fan-out gate, $F \left (\alpha \Ket{0} + \beta \Ket{1}\right ) = \alpha \Ket{00} + \beta \Ket{11}$, whenever the same data is needed by more than one proposition.  This is similar to what is done in the classical circuit model, where classical fan-out gates are used to share information between the different propositions being tested.  However, the disadvantage of this is that quantum fan-out does not have an interpretation within the standard formalism of SQL, so SQL would have to be extended to take this into account.

In this letter it was shown that an SQL proposition can be tested on a quantum computer with nonzero probability of success.  More than this cannot be expected, since a pair $\{s ,\lnot s\}$ does not generally correspond to a physically implementable test in the Hilbert space model.  This means that SQL cannot really be regarded as ``the logic of quantum computing'' in the same sense as Boolean logic is ``the logic of classical computing''.  For that, one needs to find a logic for which every quantum computation can be viewed as testing propositions in that logic and where the elementary operations of the computation correspond to the elementary connectives of the logic.

One might hope to achieve this by modifying the definition of sequential conjunction in SQL so that all pairs  $\{s ,\lnot s\}$ \emph{do} correspond to physically implementable tests.  It is unclear how this can be achieved, but there \emph{is} a natural definition of sequential exclusive OR ($\oplus_{\text{seq}}$) that always generates physical tests when combined with the $\lnot$ operator, namely $[s \oplus_{\text{seq}} t] = [\lnot t][s] + [t][\lnot s]$.  Unfortunately, exclusive OR and NOT do not constitute a universal set of gates in the classical circuit model, so they are probably not sufficient in the quantum case either, but one might hope to define a sequential conjunction satisfying a formula such as $s \oplus_{\text{seq}} t = \lnot (\lnot s \sqcap \lnot t) \sqcap \lnot(s \sqcap t)$ in analogy with the defining formula of exclusive OR in Boolean logic.

Given that SQL is not the logic of quantum computing, it is interesting to ask what model of computing it \emph{is} the logic of.  Aaronson has recently introduced a model of quantum computing with post-selection \cite{Aaronson}, wherein one is able to prescribe which outcome of a measurement that would normally be probablistic obtains with certainty.  This turns out to have interesting connections with classical complexity theory, enabling some results to be proven much more simply than by using traditional methods \footnote{In particular, the analog of BQP in this model, PostBQP, is equal to the classical complexity class PP.}.  It seems plausible that every efficient computation in this model might have an efficient description in terms of tests of SQL propositions, in which case SQL would be the logic of the model and it should give new insights into the associated complexity classes.

%Finally, an interesting possibility raised by this work is a potential connection between DMRG schemes and quantum computation.  Each stage of the algorithm presented has the same form a renormalization step in the DMRG algorithm, i.e. a local unitary operation followed by a projection of a pair of systems onto the Hilbert Space of a single system.  This is somewhat reminiscent of the connection between teleportation-based quantum computing and graph state computation in that the latter can be obtained by applying a DMRG-like transformation to the former.  In that case, one terminates the DMRG operations at one level and then switches to the single-qubit measurements prescribed by the graph state model.  The present work suggests that DMRG-like transformations could be used to effect the entire computation, with single qubit measurements only being performed to extract the output at the end of the computation.  Indeed DMRG-like operations have a lot of intuitive appeal, since they take a two-system input and produce a one system output, which is closely analogous to the classical AND gate producing a one bit output from a two bit input.  Thus, DMRG operations are natural candidates for constructing an \emph{irreversible} model of quantum computing, which is assumed to be impossible by many.  Given that the classical irreversible circuit model has a clearer logical interpretation than its reversible equivalent, such a model might have the clearest logical interpretation of any model of quantum computing proposed so far. 

Acknowledgments: I would like to thank Howard Barnum and Elham Kashefi for useful discussions and Rob Spekkens for comments on an earlier version of this manuscript.

\end{document}